\documentclass{birkjour}
\usepackage[OT1,T1]{fontenc}
\usepackage[latin1]{inputenc}
\usepackage{amsmath}
\usepackage{undertilde} 
\newcommand{\di}{\mathrm{d}} 
\newcommand{\ou}[3]{{#1}{}^{#2}{}_{#3}} 
\newcommand{\uo}[3]{{#1}{}_{#2}{}^{#3}} 
\newcommand{\I}{\mathrm{i}} 
\newcommand{\E}{\mathrm{e}} 
\newcommand{\ellp}{{\ell_{\mathrm{P}}}} 
\newcommand{\odens}{\widetilde{\eta}} 
\newcommand{\udens}{\utilde{\eta}} 
\newcommand{\CC}{\mathrm{cc.}} 
\newcommand{\eqalign}[1]{\begin{split}#1\end{split}}
\usepackage{graphicx}
\usepackage{color}

\newcommand{\qq}[1]{``#1''} 
\newcommand{\klamm}{\;\right\}} 
\usepackage{hyperref}
\begin{document}
\title[Complex Variables \& Reality Conditions for Holst's Action]{Complex Ashtekar Variables and Reality Conditions for Holst's Action}
\author{Wolfgang M. Wieland}
\address{Centre de Physique Théorique\footnote{Unité mixte de recherche (UMR 6207) du CNRS et des Universités de Provence (Aix Marseille I), de la Mediterranée (Aix-Marseille II) et du Sud (Toulon-Var); laboratoire affilié à la FRUMAM (FR 2291).},\\
   Campus de Luminy, Case 907\\
   13288 Marseille, France, EU}
\email{Wolfgang.Wieland@cpt.univ-mrs.fr}
\begin{abstract}
From the Holst action in terms of complex valued Ashtekar variables additional reality conditions mimicking the linear simplicity constraints of spin foam gravity are found. In quantum theory with the results of Ding and Rovelli we are able to implement these constraints weakly, that is in the sense of Gupta and Bleuler. The resulting kinematical Hilbert space matches the original one of loop quantum gravity, that is for real valued Ashtekar connection. Our result perfectly fit with recent developments of Rovelli and Speziale concerning Lorentz covariance within spin-form gravity.
\end{abstract}
\maketitle
\section{Introduction}
\paragraph{Motivation and Overview.} Loop quantum gravity comes in two technically different versions. On the one hand there is the canonical framework \cite{thiemann, lectures, status} trying to solve the Wheeler--DeWitt constraint equation directly. On the other hand there is the path integral approach of spin-foam gravity \cite{simplemodel, LQGvertexfinite} searching to compute transition amplitudes. Both approaches are interconnected, however the precise mathematical relation between them two is still a matter of debate. 

In a couple of papers Ding You and Carlo Rovelli \cite{physbound, covol} shed much light on this relation. They proved that within the spin-foam formalism one can actually recover the kinematics of loop quantum gravity. In this paper we are interested in something similar. The present spin-foam models are formulated in terms of $SL(2,\mathbb{C})$ holonomies. The canonical program has however much abandoned this gauge group in favour of $SU(2)$ instead. In order to build a mathematical framework allowing us to compare both formulations properly, it seems reasonable to repeat the program of canonical quantisation with respect to complex variables. 
To do this section \ref{sect1} recalls the Holst action in terms of complex Ashtekar variables. (See also the work of Alexandrov \cite{realcond} for a very similar approach towards this goal.)

Next, within section \ref{sect3} by the use of \qq{time gauge} (i.e. a partial gauge fixing aligning the internal time direction to the surface normal of the $t=\mathrm{const.}$ hyper-surfaces) we successfully introduce a Hamiltonian and perform the Dirac analysis of the constraint equations. This gauge condition manifestly breaks $SL(2,\mathbb{C})$ invariance; residual gauge transformations are restricted to the $SU(2)$ subgroup. But this does certainly not force us to write everything in terms of $SU(2)$ variables (e.g. in terms of the real valued Ashtekar connection ${}^{(\beta)}A=\Gamma+\beta K$, and its momentum conjugate). In fact the constraint equations can be split into those being $SL(2,\mathbb{C})$ gauge invariant and those breaking this symmetry. The former (i.e. the collection of Gau\ss, vector and Hamiltonian constraints) happen to be of first class, whereas the latter are of second class. The additional second class constraints, absent in the real valued formulation restrict the gauge transformations allowed to those preserving the internal time direction, and, ipso facto, break $SL(2,\mathbb{C})$ invariance. One of these \qq{reality conditions} will match the linear simplicity constraints
\begin{equation*}
C\propto K+\beta L\stackrel{!}{=}0
\end{equation*}
of spin-foam \cite{LQGvertexfinite} gravity. 

Being interested to perform the program of canonical quantisation we are then let to the crucial question whether or not the same kinematical Hilbert space as in the $SU(2)$ case emerges. Remember that this question is highly non trivial, already in quantum \emph{mechanics} when asking what happens if quantisation is performed starting from some arbitrarily chosen canonical pair. At the end of section \ref{sect4} this question can be answered in the affirmative. Starting from complex Ashtekar variables we are able to recover the kinematical Hilbert space of loop quantum gravity.

\paragraph{Notation.} In this work small indices $a,b,c,\dots$ refer to abstract indices in tangent space, internal indices in four dimensional Minkowski space are marked by capitals $I,J,K,\dots$ their three dimensional counterparts are denoted by small indices $i,j,k,\dots$ from the middle of the roman alphabet. On $\mathfrak{sl}(2,\mathbb{C})$ after choosing some internal time direction any element $X$ can be decomposed according to $X^i\tau_i$ into three \emph{complex} components $X^i\in\mathbb{C}$. Here $\tau^i:=\frac{1}{2\I}\sigma^i$ are the canonical generators of $SU(2)$ built from the Pauli spin matrices $\sigma^i$. According to this decomposition the parity transformed element $\bar{X}\in\mathfrak{sl}(2,\mathbb{C})$ is nothing but the complex conjugate $\bar{X}=\bar{X}^i\tau_i$ of the component vector $X^i\in\mathbb{C}^3$.
\section{Holst action revisited}\label{sect1}
The kinematical Hilbert space \cite{rovelli, thiemann, status} of loop quantum gravity can most naturally be obtained from the Hamiltonian formulation emerging from the so-called\footnote{In fact it is rather misleading to call it that way. Holst though proving this action to reproduce the $SU(2)$ Ashtekar variables, did actually not introduce it first. This was done by Hojman et al. \cite{Parviol} already in the 1980ies. I'm grateful to Friedrich Hehl for pointing this out.} Holst \cite{holst} action. In terms of the co-tetrad field $\eta^I$, and the $\mathfrak{so}(1,3)$-valued spin connection $\ou{\omega}{I}{J}$ it can be written according to
\begin{equation}
\eqalign{
S_{\mathrm{Holst}}[\eta,\omega] =&\frac{\hbar}{4\ellp^2}\Big[\int_M\Big(\epsilon_{IJKL}\eta^I\wedge\eta^J-\frac{2}{\beta}\eta_K\wedge\eta_L\Big)\wedge R^{KL}[\omega]+\\
&-2\int_{\partial M}\epsilon_{IJKL}\eta^I\wedge\eta^J\wedge\big(n^K\overset{\omega}{\mathcal{D}} n^L\big)\Big].\label{action}
}
\end{equation}
Here $\ou{R}{I}{J}[\omega]=\di\ou{\omega}{I}{J}+\ou{\omega}{I}{M}\wedge\ou{\omega}{M}{J}$ denotes the curvature corresponding to the spin connection, internal indices ($I,J,K,\dots\in\{0,1,2,3\}$) are raised by the flat Minkowski metric $\eta_{IJ}=\mathrm{diag}(-1,1,1,1)_{IJ}$ and $\epsilon_{IJKL}$ is the internal Levi-Civita tensor determined from $\epsilon_{0123}=1$. The last part of \eqref{action} absent in the original paper \cite{holst} is an integral over the three dimensional boundary of $M$. In the case of the Hilbert--Palatini Lagrangian this boundary term was already introduced by Obukhov \cite{obukhov}. It is built from the covariant derivative $\overset{\omega}{\mathcal{D}}=\di+\omega$ of the internal vector $n_I=\uo{\eta}{I}{a}n_{a}$ associated to the surface normal $n_a$ of $\partial M$. For an exhaustive analysis of surface terms within the Holst setting we strongly recommend \cite{surholst}. Later by passing to the Hamiltonian formulation we will implicitly assume that $\partial M$ is formed by two Cauchy surfaces $\Sigma_0$ and $\Sigma_1$ glued at spatial infinity. Several constants appear in this action; $\ellp=\sqrt{8\pi\hbar G/c^3}\approx 8\cdot 10^{-35}\mathrm{m}$ equals the rescaled Planck length, and the Barbero--Immirzi parameter $\beta$ is a dimensionless parameter unique to all models of loop quantum gravity. Leaving classical dynamics of general relativity unaffected this number can affect quantum theory only.

In order to derive the equations of motion for general relativity from the principle of least action appropriate boundary conditions have to be imposed. Otherwise the variation principle remains obscure. In the case of the Holst action we will only allow for variations $\delta\eta^I$ and $\delta\omega^{IJ}$ of the elementary configuration variables subject to the following restrictions on the boundary
\begin{eqnarray}
\ou{h}{I}{J}\,\big(\varphi^\ast\delta\eta^J\big)\stackrel{!}{=}0,\label{conda}\\
\ou{h}{I}{M}\ou{h}{J}{N}\,\big(\varphi^\ast\delta\omega^{MN}\big)\stackrel{!}{=}0.\label{condb}
\end{eqnarray}
Here the internal projector $\ou{h}{I}{J}$ annihilating $n^I$ together with the embedding $\varphi:\partial M\rightarrow M$ and the corresponding pull-back $\varphi^\ast$ have been used.

We are now interested to decompose the Holst action defined by \eqref{action} into its self- and antiselfdual parts
\begin{equation}
S_{\mathrm{Holst}}[\eta,\omega] = -\frac{\hbar}{2\ellp^2}\frac{\beta+\I}{\I\beta}S_{\mathbb{C}}+\frac{\hbar}{2\ellp^2}\frac{\beta-\I}{\I\beta}\bar{S}_{\mathbb{C}}+I_{\partial{M}}.
\end{equation}
Here we have introduced the $\mathbb{C}$-valued action
\begin{equation}
S_{\mathbb{C}}=\int_MP_{IJMN}\eta^I\wedge\eta^J\wedge R^{MN}[\omega],
\end{equation}
together with the selfdual projector
\begin{equation}
\ou{P}{IJ}{MN}:=\frac{1}{2}\big(\delta^{I}_{[M}\delta^J_{N]}-\frac{\I}{2}\ou{\epsilon}{IJ}{MN}\big),
\end{equation}
fulfilling the elementary properties:
\begin{equation*}
\eqalign{
\mbox{(i)}\;	 &\ou{P}{IJ}{MN}\ou{P}{MN}{LK}=\ou{P}{IJ}{LK},\\
\mbox{(ii)}\;  &P_{IJMN}=-P_{JIMN}=-P_{IJNM}=P_{MNIJ},\\
\mbox{(iii)}\; &\ou{\bar{P}}{IJ}{MN}\ou{P}{MN}{LK}=0,\\
\mbox{(iv)}\;  &\ou{\epsilon}{IJ}{MN}\ou{P}{MN}{LK}=2\I\ou{P}{IJ}{LK}.
}
\end{equation*}
See also \cite{realcond} on that. Notice that the boundary term represented by $I_{\partial{M}}$ has not been decomposed into self- and antiselfdual parts. 

In order to construct a Hamiltonian formulation of the theory defined by \eqref{action} let us first decompose all four-dimensional quantities into their  spatio-temporal components. This is achieved \cite{wald} by introducing a global time function $t:M\rightarrow \mathbb{R}$ (thereby implicitly assuming global hyperbolicity, i.e. $M=(0,1)\times\Sigma$) together with a future directed vector field $t^a$ transversal to all $t=\mathrm{const.}$ hyper-surfaces $\Sigma_t$ (i.e. $t^a\partial_at=1$). We can then introduce the spatial components of the elementary configuration variables, namely
\begin{eqnarray}
 \mbox{co-triad:} &e^i &\hspace{-0.3cm}= \mathrm{pr}\,\eta^i,\\
 \mbox{$\mathfrak{so}(3)$-connection:} &\ou{\Gamma}{i}{j} &\hspace{-0.3cm}=   
  \mbox{pr}\,\ou{\omega}{i}{j},\\
 \mbox{extrinsic curvature:} & K^i &\hspace{-0.3cm}=\mathrm{pr}\,\ou{\omega}{i}{o}.
\end{eqnarray}
Here we have used the spatial projection $\mathrm{pr}\,\varphi:=\varphi-\di t\wedge\iota_t\varphi$ of any four-dimensional $p$-form $\varphi$ onto $\Sigma_t$.
Still additional variables representing the $\di t$ components are missing. Introducing lapse-function $N$, shift-vector $N^a$ together with additional \qq{Lagrangian multiplier fields} $\ou{\phi}{i}{o}$ and $\ou{\phi}{i}{j}$ representing both infinitesimal boosts and rotations along the time axis, we are left with the following decomposition of the four-dimensional configuration variables 
\begin{align}
\eta^o=N\di t,&\qquad\eta^i=N^a\ou{e}{i}{a}\di t+e^i,\label{decom1}\\
\ou{\omega}{i}{o}=\di t\ou{\phi}{i}{o}+K^i,&\qquad\ou{\omega}{i}{j}=\di t\ou{\phi}{i}{j}+\ou{\Gamma}{i}{j}.\label{decom2}
\end{align}
Notice that local Lorentz invariance allowed us to choose \qq{time gauge} thereby setting $\mathrm{pr}\,\eta^o$ globally to zero. Inserting this decomposition of variables into the selfdual action we recover the original expression of Ashtekar \cite{ashtekar}
\begin{equation}
\eqalign{S_{\mathbb{C}}=\int_{\mathbb{R}}\di t\int_\Sigma&-\uo{E}{i}{a}(\ou{\mathcal{L}_t{A}}{i}{a}-D_a\Lambda^i)+N^a\ou{F}{i}{ab}\uo{E}{i}{b}+\\
&+\frac{\I}{2}\utilde{N}\uo{\epsilon}{i}{lm}\ou{F}{i}{ab}\uo{E}{l}{a}\uo{E}{m}{b}.}\label{cactn}
\end{equation}
Where we have just introduced the old i.e. complex valued Ashtekar \cite{newvariables} variables 
\begin{eqnarray}
\mbox{Ashtekar connection:}&&
\hspace{-0.3cm}\ou{A}{i}{a}=\ou{\Gamma}{i}{a}+\I\ou{K}{i}{a},\\
\mbox{Densitised triad:}&&\hspace{-0.3cm}\uo{E}{i}{a}=\frac{1}{2}\odens^{abc}\epsilon_{ilm}\ou{e}{l}{b}\ou{e}{m}{c}.
\end{eqnarray}
In equation \eqref{cactn} $\mathcal{L}_t$ equals the Lie derivative along the time-flow vector field, $\odens^{abc}$ is the spatial Levi-Civita tensor density, $\mathfrak{so}(3)$ elements have been decomposed into the generators of $SO(3)$ according to e.g. $\ou{\Gamma}{i}{j}=\ou{\epsilon}{i}{mj}\Gamma^m$, the Ashtekar covariant derivative is denoted by $D_a=\partial_a+[A_a,\cdot]$, $\ou{F}{i}{ab}=[D_a,D_b]^i$ is the curvature associated, and $\Lambda^i:=\phi^i+\I\ou{\phi}{i}{o}$. Furthermore $\utilde{N}:=\mu^{-1}N$ is a density of weight minus one, where $\mu=e^1\wedge e^2\wedge e^3$ represents the oriented volume element on $\Sigma$, and $\mu^{-1}=\uo{e}{1}{\mu}\uo{e}{2}{\nu}\uo{e}{3}{\rho}\partial_\mu\wedge\partial_\nu\wedge\partial_\rho$ equals its inverse.

What about the surface contribution? In terms of the decomposition introduced above $I_{\partial M}$ takes the usual form of the Gibbons--Hawking--York \cite{hawking, york} boundary term
\begin{equation}
\!I_{\partial M}=-\frac{\hbar}{\ellp^2}\int_{\partial M}\uo{E}{i}{a}\ou{K}{i}{a}=-\frac{\hbar}{\ellp^2}\int_{\Sigma_1}\uo{E}{i}{a}\ou{K}{i}{a}+\frac{\hbar}{\ellp^2}\int_{\Sigma_0}\uo{E}{i}{a}\ou{K}{i}{a}.
\end{equation}
By the use of time-gauge the boundary conditions (\ref{conda}, \ref{condb}) on $\partial M$ simplify
\begin{equation}
\delta\uo{E}{i}{a}|_{\partial{M}}=0,\qquad\delta\ou{\Gamma}{i}{a}|_{\partial{M}}=0.
\end{equation}
However there are no restrictions on the variations of lapse $N$, shift $N^a$ and extrinsic curvature $\ou{K}{i}{a}$ on the boundary.
\section{Hamiltonian formulation}\label{sect3}
\subsection{Phase space, constraints and time evolution}
As a matter of fact Hamiltonian mechanics is about time evolution. In particular one tries to split the equations of motion into two parts. First of all one has the evolution equations generated by the Hamiltonian vector field $X_{H}=\{H,\cdot\}$ describing dynamics of the theory (e.g. the evolution equation for the electric field $\partial_t\vec{E}=\vec{\nabla}\times(\vec{\nabla}\times\vec{A})$ in terms of the vector potential). Secondly one might have additional constraint equations discarding all unphysical degrees of freedom (e.g. the Gau\ss-law $\vec{\nabla}\cdot\vec{E}=0$).

Furthermore the Hamiltonian formulation is not necessarily about performing a Legendre transformation. In particular the decomposition of the complex action \eqref{cactn} strongly suggests not to do so. The action is already in Hamiltonian form, performing a singular Legendre transformation would introduce an unnaturally large phase space containing momenta associated to densitised lapse $\utilde{N}$, shift vector, to $\Lambda^i$ and to both the densitised triad and the connection. But this is not needed at all.

In order to turn the equations of motion into Hamiltonian form let us first introduce the \qq{natural} phase space $\mathcal{P}$ of smooth $SL(2,\mathbb{C})$ connections $\ou{A}{i}{a}$ and corresponding momenta $\uo{\Pi}{i}{a}$. The symplectic structure is defined by the only non vanishing Poisson brackets, namely
\begin{equation}
\eqalign{&\big\{\uo{\Pi}{i}{a}(p),\ou{A}{j}{b}(q)\big\}=\delta^j_i\delta^a_b\delta^{(3)}(p,q),\\
&\big\{\uo{\bar{\Pi}}{i}{a}(p),\ou{\bar{A}}{j}{b}(q)\big\}=\delta^j_i\delta^a_b\delta^{(3)}(p,q).}\label{poissklamm}
\end{equation}
Notice that $\uo{\Pi}{i}{a}\otimes\tau^i$ has to be understood as $\mathfrak{sl}(2,\mathbb{C})$ valued vector density, $\delta^{(3)}(p,q)$ is the Dirac distribution (a scalar density) and $\bar{X}$ denotes complex conjugation of $X$. 

Look at the first part of \eqref{cactn}, which actually tell us that $\uo{\Pi}{i}{a}$ and $\uo{\bar{\Pi}}{i}{a}$ are related according to
\begin{align}
\uo{\Pi}{i}{a}\big|_{\mathrm{EOM}}&=+\frac{\hbar}{2\ellp^2}\frac{\beta+\I}{\I\beta}\uo{E}{i}{a},\label{cmpxconsa}\\
\uo{\bar{\Pi}}{i}{a}\big|_{\mathrm{EOM}}&=-\frac{\hbar}{2\ellp^2}\frac{\beta-\I}{\I\beta}\uo{E}{i}{a}.\label{cmpxconsb}
\end{align}
And the abbreviation EOM should remind us that these relations have to be fulfilled when the equations of motion hold. However we can put equations (\ref{cmpxconsa}, \ref{cmpxconsb}) upside down, allowing us to define $\uo{E}{i}{a}$ on the entire phase space
\begin{equation}
\eqalign{
\uo{E}{i}{a}:\!&=\frac{\ellp^2}{\hbar}\Big(\frac{\I\beta}{\beta+\I}\uo{\Pi}{i}{a}-\frac{\I\beta}{\beta-\I}\uo{\bar{\Pi}}{i}{a}\Big)=\\
&=\frac{\ellp^2}{\hbar}\frac{\beta}{\beta^2+1}\Big((\uo{\Pi}{i}{a}+\uo{\bar{\Pi}}{i}{a})+\I\beta(\uo{\Pi}{i}{a}-\uo{\bar{\Pi}}{i}{a})\Big).}
\end{equation}
Similarly equations (\ref{cmpxconsa}, \ref{cmpxconsb}) tell us that in order to recover the original number of degrees of freedom the quantity
\begin{equation}
\eqalign{
\uo{C}{i}{a}:\!&=\frac{\ellp^2}{\I\hbar}\Big(\frac{\I\beta}{\beta+\I}\uo{\Pi}{i}{a}+\frac{\I\beta}{\beta-\I}\uo{\bar{\Pi}}{i}{a}\Big)=\\
&=\frac{\ellp^2}{\hbar}\frac{\beta}{\beta^2+1}\Big(-\I(\uo{\Pi}{i}{a}-\uo{\bar{\Pi}}{i}{a})+\beta(\uo{\Pi}{i}{a}+\uo{\bar{\Pi}}{i}{a})\Big)\stackrel{\mathrm{EOM}}{=}0}\label{ccons}
\end{equation}
is constrained to vanish. Due to the striking similarity with the constraints $\uo{C}{i}{a}=\uo{E}{i}{a}-\uo{\bar{E}}{i}{a}=0$ found within the old Ashtekar approach we call them reality conditions. Let us stop here for a moment. Equation \eqref{ccons} strongly suggests to introduce real and imaginary parts of ${\Pi}$ corresponding to both internal rotations and boosts; i.e. we set $L\propto\Pi+\bar{\Pi}$, and $\I K\propto\Pi-\bar\Pi$. Dropping all decorating indices the reality conditions turn into
\begin{equation}
\boxed{C\propto K+\beta L =0.}
\end{equation}
Hence $C=0$ though formal at this level takes the basic form of the linear \cite{LQGvertexfinite, flppdspinfoam} simplicity constraints of spin foam gravity \cite{simplemodel}. Within section \ref{classal} this relation will become more explicit. Perez and Rezende \cite{Holstplustop} have recovered the very same constraints from a more general setting, i.e. the Holst action augmented with all possible topological invariants together with a term containing the cosmological constant.

\noindent In order to proceed let us define the following smeared quantities
\begin{align}
 \mbox{reality condition:}&\quad\uo{C}{i}{a}[\ou{V}{i}{a}]  
	:=\int_{\Sigma}\uo{C}{i}{a}\ou{V}{i}{a}\stackrel{\mathrm{EOM}}{=}0,
	\label{real}\\
 \mbox{Gau\ss\;constraint:}&\quad G_i[\Lambda^i] :=  
	\int_\Sigma \Big(-\Lambda^iD_a\uo{\Pi}{i}{a}+\CC\Big)
	\stackrel{\mathrm{EOM}}{=}0, \label{gauss}\\
 \mbox{vector constraint:}&\quad H_a[N^a] :=
	\int_\Sigma\Big(N^a\ou{F}{i}{ab}\uo{\Pi}{i}{b}+\CC\Big)
	\stackrel{\mathrm{EOM}}{=}0,\label{vec} \\
 \begin{split}\mbox{Hamiltonian constraint:}&\quad H[\utilde{N}] :=
  -\frac{\ellp^2}{\hbar}\int_\Sigma\utilde{N}\Big(\frac{\beta}{\beta+\I}\cdot\\
	&\qquad\!
	\cdot\uo{\epsilon}{i}{lm}\ou{F}{i}{ab}\uo{\Pi}{l}{a}\uo{\Pi}{m}{b}+\CC\Big)
	\stackrel{\mathrm{EOM}}{=}0,\label{ham}
 \end{split}
\end{align}
where the symbol $\CC$ means complex conjugation of everything preceding. Variation of the original action (\ref{action}, \ref{cactn}) with respect to lapse, shift-vector and $\Lambda^i$ immediately reveals that all of these quantities are constrained to vanish provided the equations of motion hold. 

The physical meaning of both Gau\ss\;and vector constraint is immediate. The (Hamiltonian vector field of the) Gau\ss\;constraint $G_i[\Lambda^i]$ generates $SL(2,\mathbb{C})$ gauge transformations  $g_\Lambda=\exp(\Lambda^i\tau_i)$ on phase space:
\begin{align}
  \exp\big(X_{G_i[\Lambda^i]}\big)\Pi^a  & = 
	g_\Lambda^{-1}\Pi^ag_\Lambda,\\
  \exp\big(X_{G_i[\Lambda^i]}\big)A_a & = 
	g_\Lambda^{-1}\partial_a g_\Lambda+g_\Lambda^{-1}A_ag_\Lambda.
\end{align}
Thus the momentum $\Pi^a\equiv\uo{\Pi}{i}{a}\otimes\tau^i$ transforms under the adjoint representation of $SL(2,\mathbb{C})$. Except for the reality condition $\uo{C}{i}{a}=0$ all constraints introduced above are actually invariant under these gauge transformations. The vector constraint on the other hand generates spatial diffeomorphisms modulo $SL(2,\mathbb{C})$ gauge transformations. In both cases the proof follows the lines of the $SU(2)$ case.

The set of constraint equations gives us the first part of the equations of motion, what about the other part, what about time evolution on phase space? In order to study the evolution equations let us first introduce Dirac's \emph{primary Hamiltonian} (see for instance \cite{dirac} and also \cite{thiemann} for the details of the following formalism)
\begin{equation}
H^\prime:=\uo{C}{i}{a}[\ou{V}{i}{a}]+G_i[\Lambda^i]+H_a[N^a]+H[\utilde{N}].\label{primham}
\end{equation}
Notice the appearance of a $\uo{C}{i}{a}[\ou{V}{i}{a}]$ term proportional to the reality conditions \eqref{ccons}. We will later comment on the necessity of this additional expression absent in the original action \eqref{cactn}.

For any functional $X$ of the phase space variables, let us first \emph{define} its time evolution according to
\begin{equation}
\mathcal{L}_tX=\{H^\prime,X\}.\label{evolv}
\end{equation}
Having introduced the symplectic structure \eqref{poissklamm} essentially by hand, it is still an open question whether or not Hamiltonian time evolution defined in the sense of \eqref{evolv} is compatible with the Euler--Lagrange equations of motion. In order to check this, one first proves equivalence between the evolution equations derived from the variation principle and the following Hamiltonian equations
\begin{align}
\Big(\frac{\beta+\I}{\I\beta}\mathcal{L}_t\ou{A}{i}{a}+\CC\Big)\Big|_{C=0}& = 
\Big\{H^\prime,\frac{\beta+\I}{\I\beta}\ou{A}{i}{a}+\CC\Big\}\Big|_{C=0},\label{avelos1}\\
\big(\mathcal{L}_t\uo{E}{i}{a}\big)\big|_{C=0} & =
\big\{H^\prime,\uo{E}{i}{a}\big\}\big|_{C=0}.\label{evelos}
\end{align}
And $C=0$ is an abbreviation for $\uo{C}{i}{a}(p)=0$. Notice that the Euler--Lagrangian evolution equations do not determine $\mathcal{L}_tA$ and $\mathcal{L}_t\bar{A}$ independently, but only the linear combination appearing in \eqref{avelos1}. What about the other linearly independent combination of $A$ and $\bar{A}$? Calculating the corresponding Poisson bracket reveals
\begin{equation}
\eqalign{\Big(\frac{\beta+\I}{\beta}\mathcal{L}_t\ou{A}{i}{a}&+\CC\Big)\Big|_{C=0}  = \frac{2\ellp^2}{\hbar}\ou{V}{i}{a}+\\
&+\Big\{G_i[\Lambda^i]+H_a[V^a]+H[\utilde{N}],\frac{\beta+\I}{\beta}\ou{A}{i}{a}+\CC\Big\}\Big|_{C=0}\label{avelos2}.}
\end{equation}
 As a matter of fact the left hand side of this equation, i.e. the time derivative of $\ou{\Gamma}{i}{a}-\frac{1}{\beta}\ou{K}{i}{a}$ does not appear in the list of evolution equations derived from the Euler--Lagrangian framework. But still, this quantity is fully determined by all the equations of motion: Variation of the action with respect to the $\mathfrak{so}(1,3)$ spin connection immediately reveals that torsion is forced to vanish. But if there is no torsion the evolution equations \eqref{avelos1} and \eqref{evelos} fully determine the left hand side of \eqref{avelos2}.

All of this happens in the case of the Lagrangian formulation, but in the Hamiltonian framework the situation is slightly different. The vanishing of torsion emerges only in a secondary step explicitly studied later. Unable to use this constraint in order to calculate the left hand side of \eqref{avelos2} we fix it by an additional Lagrangian multiplier $\ou{V}{i}{a}$. This multiplier determines the time derivative \eqref{avelos2} to some yet unspecified value $\uo{V}{i}{a}+\dots$, but leaves both \eqref{avelos1} and \eqref{evelos} unchanged. Its value, later found to be $\uo{V}{i}{a}=0$, is derived by the requirement that all constraints are preserved under the time evolution generated by \eqref{evolv}. All of this is naturally achieved by the choice of the primary Hamiltonian \eqref{primham} introduced above. 

We have already proven \eqref{evelos} that the Hamiltonian time evolution for $\uo{E}{i}{a}$ is perfectly consistent with the equations of motion found from the variation principle, but what about $\uo{C}{i}{a}$, i.e. the other linearly independent combination of $\uo{\Pi}{i}{a}$ and $\uo{\bar\Pi}{i}{a}$? Comparison with the Lagrangian framework reveals that the quantity 
\begin{equation}
\mathcal{L}_t\uo{C}{i}{a}\big|_{C=0}=\big\{H^\prime,\uo{C}{i}{a}\big\}\big|_{C=0}\stackrel{\mathrm{EOM}}{=}0.
\end{equation}
is actually constrained to vanish. 
Which of course is nothing but the statement that the reality conditions have to be preserved under time evolution on phase space. This finishes the proof of compatibility between time evolution generated by \eqref{evolv} and the corresponding  Euler--Lagrangian evolution equations.
\subsection{Dirac consistency of the constraints}
In this section we check if all the constraints (\ref{real}, \ref{gauss}, \ref{vec}, \ref{vec}, \ref{ham}) are preserved under the time evolution defined by \eqref{evolv}. As a preliminary step to calculate this one first proves the following list of Poisson brackets:
\begin{align}
&\big\{G_i[\Lambda^i],G_j[M^j]\big\}  = -G_i\big[[\Lambda,M]^i\big]\label{ggbrace}\\
&\big\{G_i[\Lambda^i],H_a[V^a]\big\}  =  0\label{gvbrace}\\
&\big\{G_i[\Lambda^i],H[\utilde{N}]\big\}  =  0\label{ghbrace}\\
&\big\{H_a[U^a],H_b[V^b]\big\}  =  -H_a\big[[U,V]^a\big]-G_i\big[F^i(U,V)\big]\label{vvbrace}\\
&\big\{H_a[V^a],H[\utilde{N}]\big\}  = -H\big[\mathcal{L}_V\utilde{N}]-G_i\Big[\frac{\delta H[\utilde{N}]}{\delta \uo{\Pi}{i}{a}}V^a\Big]\label{vhbrace}\\
\begin{split}&\big\{H[\utilde{M}],H[\utilde{N}]\big\}  = \frac{4\ellp^4}{\hbar^2}\int_\Sigma\frac{\beta^2}{(\beta+\I)^2}
(\utilde{M}\partial_c\utilde{N}-\utilde{N}\partial_c\utilde{M})\uo{\Pi}{j}{c}\Pi^{{j}{a}}\cdot\\
&\qquad\qquad\qquad\qquad\qquad\qquad\cdot\ou{F}{i}{ab}\uo{\Pi}{i}{b}+\CC\label{hhbrace}
\end{split}
\end{align}
Here $[U,V]=\mathcal{L}_UV$ is the Lie bracket between vector fields, $[\Lambda,M]^i:=\ou{\epsilon}{i}{lm}\Lambda^lM^m$ is the commutator on $\mathfrak{sl}(2,\mathbb{C})$, and $\mathcal{L}_V\utilde{N}=-\utilde{N}^2\partial_a(V^a\utilde{N}^{-1})$ is the Lie derivative of the inverse density. Observe that the Poisson bracket between to smeared Hamiltonian constraints does not give a linear combination of smeared Gau\ss, vector and Hamiltonian constraints again. Therefore on the \emph{full} phase space of complex valued connections and corresponding momenta the algebra generated by $G_i[\Lambda^i]$, $H_a[N^a]$ and $H[\utilde{N}]$ does not close. Instead the Poisson bracket between two smeared Hamiltonian constraints vanishes \emph{on-shell}. This can be seen as follows. Restricting the result of \eqref{hhbrace} to those parts of phase space where $\uo{C}{i}{a}=0$ holds we find
\begin{equation}
{\big\{H[\utilde{M}],H[\utilde{N}]\big\}\Big|_{C=0}=-H_a\big[\uo{E}{j}{a}E^{jb}(\utilde{M}\partial_b\utilde{N}-\utilde{N}\partial_b\utilde{M})\big]\Big|_{C=0}.}
\end{equation}
And the right hand side of this equation being proportional to the vector constraint again vanishes on the constraint hyper-surface. Notice also that even though $\utilde{M}\partial_b\utilde{N}$ is ill defined without choosing a preferred derivative acting on the density weight (not necessarily the metric compatible one), the antisymmetric part $\utilde{M}\partial_b\utilde{N}-\utilde{N}\partial_b\utilde{M}$ perfectly is. This follows from the fact that derivatives \qq{acting} on the density weight cancel.
\paragraph{Consistency of $\boldsymbol{\uo{C}{i}{a}[\ou{V}{i}{a}]=0}$.} 
Calculating the Poisson bracket generating the corresponding time derivative reveals that
\begin{equation}
\left.\eqalign{
&\mathcal{L}_t\uo{C}{i}{a}\big|_{C=0} = \big\{ H^\prime,\uo{C}{i}{a}\big\}\big|_{C=0}=
-\frac{1}{2\I}\uo{\epsilon}{il}{m}(\Lambda^l-\bar{\Lambda}^l)\uo{E}{m}{a}+\\
&\,+\frac{1}{2\I}(D_b-\bar{D}_b)\big(N^b\uo{E}{i}{a}-N^a\uo{E}{i}{b}\big)
+\frac{1}{2}\uo{\epsilon}{i}{lm}(D_b+\bar{D}_b)\big(\utilde{N}\uo{E}{l}{b}\uo{E}{m}{a}\big)\\
& = -\uo{\epsilon}{il}{m}\Big(\frac{1}{2\I}(\Lambda^l-\bar{\Lambda}^l)-N^b\ou{K}{l}{b}-e^{lb}\partial_bN\Big)\uo{E}{m}{a}+N\odens^{bac}\nabla_be_{ic}
.}\right\}\label{realcons}
\end{equation}
Where we have reintroduced the scalar (i.e. the \qq{undensitised} lapse) $N=\mu\utilde{N}$ and $\nabla_a=\frac{1}{2}(D_a+\bar{D}_a)=\partial_a+[\Gamma_a,\cdot]$ defines another covariant derivative. In the last line of this equation both the expression within the big bracket and the $\nabla e$ term must vanish independently. This can be seen as follows. Notice that the covariant derivative $\nabla_a$ equals the Levi-Civita derivative modulo a difference tensor $\ou{\Delta}{i}{a}$. Using the fact that $\nabla_a\uo{E}{j}{a}=0$ provided both the Gau\ss\;constraint and the reality condition $\uo{C}{i}{a}=0$ hold, it follows that $\ou{\Delta}{i}{a}$ must be symmetric (in the sense of $\epsilon_{jlm}\ou{\Delta}{l}{b}E^{mb}=0$). But $\mathcal{L}_t\uo{C}{i}{a}$ must vanish, which reveals that
\begin{equation}
\eqalign{
&-\frac{1}{2\I}(\Lambda^j-\bar{\Lambda}^j)+N^a\ou{K}{j}{a}+e^{ja}\partial_aN\Big|_{C,G=0}\stackrel{!}{=}\\
&\;=-\frac{1}{2}N(e^{ic}e^{jb}-e^{ib}e^{jc})\nabla_be_{ic}\Big|_{C,G=0}=\\
&\;=-\frac{1}{2}N(e^{ic}e^{jb}-e^{ib}e^{jc})\epsilon_{ilm}\ou{\Delta}{l}{b}\ou{e}{m}{c}\Big|_{C,G=0}=
\frac{1}{2}N\uo{\epsilon}{il}{j}\ou{\Delta}{l}{b}e^{ib}\Big|_{C,G=0}=0.\label{boostpart}
}
\end{equation}
Therefore the imaginary part of $\Lambda^i$ (describing boosts along the flow of time) is completely fixed by the dynamics of the theory to the value
\begin{equation}
\frac{1}{2\I}\big(\Lambda^i-\bar{\Lambda}^i\big)=+N^a\ou{K}{i}{a}+e^{ia}\partial_aN.
\end{equation}
Since the condition $\uo{C}{i}{a}=0$ is invariant under internal rotations but does not remain valid if boosted, this restriction should not surprise us. 

Comparison with \eqref{boostpart} reveals that $\nabla e$ must vanish too. Therefore there is the additional secondary constraint forcing the difference tensor $\ou{\Delta}{i}{a}$ to vanish:
\begin{equation}
\ou{\Delta}{i}{a}=\frac{1}{2}\big(\ou{A}{i}{a}-\ou{\bar{A}}{i}{a}\big)-\ou{\overset{\mathrm{LC}}{\Gamma}}{i}{a}[E]\stackrel{\mathrm{!}}{=}0.\label{difftens}
\end{equation}
Where $\ou{\overset{\mathrm{LC}}{\Gamma}}{i}{a}[E]$ denotes the Levi-Civita connection functionally depending on $\uo{E}{i}{a}$. This equation is highly non polynomial \cite{thiemann} in $E$, the equivalent but technically different version
\begin{equation}
2T:=D e+\bar{D}  e\stackrel{!}{=}0\label{tordef}
\end{equation}
is much simpler to handle. The latter just sets the spatial part of the four dimensional torsion 2-form to zero.
\paragraph{Consistency of $\boldsymbol{T=0}$.}
Here we just mention the final result, i.e:
\begin{equation}
\eqalign{
\mathcal{L}_t\ou{T}{i}{ab}\big|_{C,G,T=0}&=\frac{1}{2}
\big\{H^\prime,\ou{(D e)}{i}{ab}+\CC\big\}\big|_{C,G,T=0}=\\
&=\frac{1}{2}\ou{\epsilon}{i}{lm}\big(\ou{V}{l}{a}\ou{e}{m}{b}-\ou{V}{l}{b}\ou{e}{m}{a}\big)\stackrel{!}{=}0.\label{tor}
}
\end{equation}
And $C,G,T=0$ shall remind us that these equations hold provided torsion $T$ vanishes and both reality condition and Gau\ss\; constraint are satisfied. We conclude by stating that equation \eqref{tor} restricts the Lagrangian multiplier $\ou{V}{i}{a}$ to the value
\begin{equation}
\ou{V}{i}{a}=0
\end{equation}
being proven by an elementary algebraic manipulation of \eqref{tor}.
\paragraph{Consistency of Gau\ss, vector and Hamiltonian constraint.}
Since the conservation of the secondary constraint $T=0$ forces the Lagrangian multiplier $\ou{V}{i}{a}$ to vanish, consistency of all remaining constraints follow already from the relations (\ref{ggbrace}, \ref{gvbrace}, \ref{ghbrace}, \ref{vvbrace}, \ref{vhbrace}, \ref{hhbrace}) presented above.

\section{Towards quantum theory}\label{sect4}
\subsection{Classically smeared algebra}\label{classal}
Searching for a quantisation of the classical phase space one realises that the Poisson brackets \eqref{poissklamm} between the elementary phase space variables behave too singular, and do not allow for direct quantisation. One therefore searches for a suitable smeared algebra, still rich enough in order to allow for reconstruction of the full phase space. This naturally turns out to be the $SL(2,\mathbb{C})$ analog of the holonomy flux algebra explicitly constructed in \cite{thiemann} for the $SU(2)$ case. 
This algebra lives on the collection of all possible graphs $\Gamma$. Where any graph consists of a finite collection of ordered oriented (piecewise analytic) paths $(\gamma_1,\dots,\gamma_L)=:\Gamma$. To each of these \qq{links} $\gamma_i$ there is a dual surface $f_i$ associated. Since $\Sigma$ is orientable each of these \qq{faces} inherits a natural orientation from $\gamma_i$.

\begin{figure}[h]
     \centering
     \includegraphics[width= 0.25\textwidth]{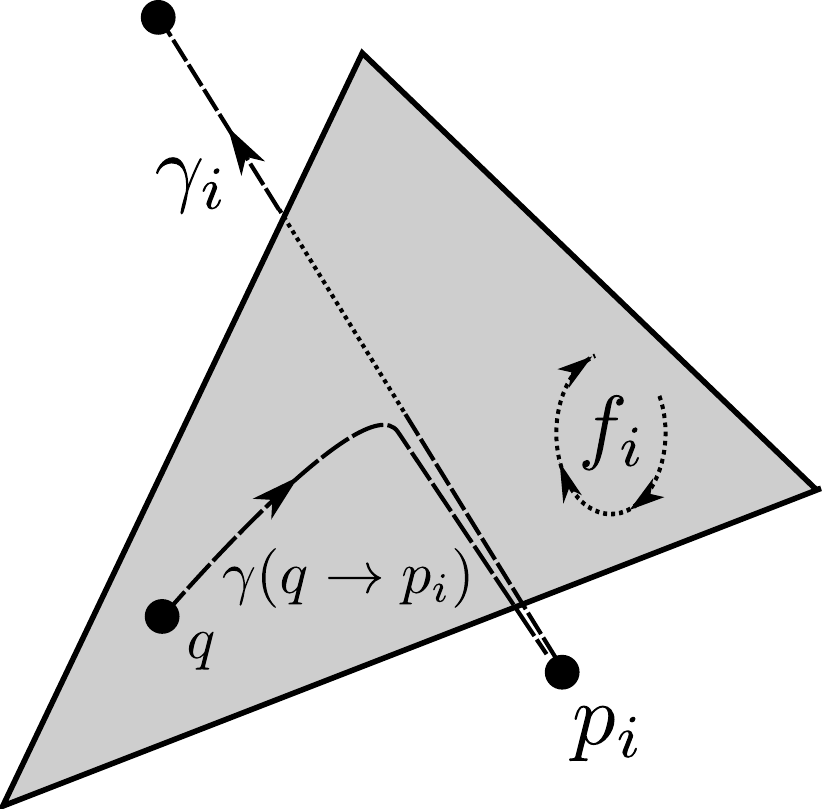}
     \caption{Phase space variables are smeared over links and faces.}
     \label{smearng}
\end{figure}

The elementary configuration variable $\ou{A}{i}{a}$ being a $SL(2,\mathbb{C})$ connection can now naturally be smeared over any of these links, thereby obtaining the holonomy or parallel propagator
\begin{equation}
U[f_i]\equiv U_{\gamma_i}=\boldsymbol{P}\exp\left(-\int_{\gamma_i}A\right)\in SL(2,\mathbb{C}),
\end{equation}
where $\boldsymbol{P}$ is the usual path-ordering symbol. Furthermore by 
\begin{equation}
\Pi[f_i]=\int_{q\in f_i}U_{\gamma(q\rightarrow p_i)}\udens_{abc}\Pi^a\big|_qU_{\gamma(q\rightarrow p_i)}^{-1}\label{fluxsmear}
\end{equation}
we obtain a natural smearing of the momentum variable over the faces dual to the links. Here $\udens_{abc}$ denotes the inverse Levi-Civita density, and therefore $\udens_{abc}\uo{\Pi}{i}{a}$ defines a 2-form which can consistently be integrated over any surface. Moreover ${\gamma(q\rightarrow p_i)}$ is a suitable family\footnote{More precisely \cite{EEcomm} the system of paths $\gamma(q\rightarrow p_i)=\gamma(p\rightarrow p_i)\circ\gamma(q\rightarrow p)$ consist of two parts; the first one, i.e. $\gamma(q\rightarrow p)$ lies within the surface $f_i$ mapping $q$ to the intersection $p=\gamma_i\cap f_i$. The second part goes from $p$ along $\gamma_i$ towards the initial point $p_i=\gamma_i(0)$.} 
of paths parallely transporting any $q$ towards the initial point $p_i=\gamma_i(0)$. The details determining this family can be found e.g. in \cite{EEcomm}, roughly sketched in figure \ref{smearng} presented bellow.
These smeared variables obey the standard Poisson algebra \cite{thiemann, steppingout, EEcomm} of lattice gauge theory, i.e. the holonomy flux algebra of $SL(2,\mathbb{C})$, that is:
\begin{align}
\big\{U[f],U[f^\prime]\big\}&=0\label{UUbrace},\\
\big\{\Pi_i[f],U[f^\prime]\big\}&=-\epsilon(f,f^\prime)
\begin{cases}U[f]\tau_i,&\mathrm{if\;}\epsilon(f,f^\prime)=+1\\\tau_iU[f],&\mathrm{if\;}\epsilon(f,f^\prime)=-1\end{cases},\label{PiUbrace}\\
\big\{\Pi_i[f],\Pi_j[f^\prime]\big\}&=-\delta_{ff^\prime}\uo{\epsilon}{ij}{k}\Pi_k[f].\label{PiPibrace}
\end{align}
Here $\epsilon(f,f^\prime)$ denotes the relative orientation of the two faces, and $\delta_{ff^\prime}=|\epsilon(f,f^\prime)|$. Notice furthermore that these Poisson brackets remain qualitatively unchanged if evaluated in some irreducible unitary representation of $SL(2,\mathbb{C})$, e.g:
\begin{equation}
\big\{\Pi_i[f],D^{(\rho,j_o)}\big(U[f]\big)\big\}=-D^{(\rho,j_o)}\big(U[f]\big)\,\ou{D}{(\rho,j_o)}{\!\!\ast}(\tau_i).\label{reppoiss}
\end{equation}
Here $D^{(\rho,j_o)}$ denote the unitary irreducible representations of $SL(2,\mathbb{C})$, and $\ou{D}{(\rho,j_o)}{\!\!\ast}$ is the corresponding induced representation of the Lie algebra $\mathfrak{sl}(2,\mathbb{C})$. Equation \eqref{reppoiss} follows from nothing but $D^{(\rho,j_o)}(\exp\omega)=\exp(\ou{D}{(\rho,j_o)}{\!\!\ast}\omega)$ for all $\omega\in\mathfrak{sl}(2,\mathbb{C})$. Furthermore $\rho\in\mathbb{R}$ is a continuous label whereas $2j_o\in\mathbb{N}_0$ is discrete, and together they fully characterise the unitary irreducible representations of the group. Any of these $SL(2,\mathbb{C})$ representations induce a representation of the $SU(2)$ subgroup. The corresponding Clebsch--Gordan decomposition leads to an infinite direct sum $D^{(j_o)}\oplus D^{(j_o+1)}\oplus\dots$ of irreducible $SU(2)$ representations $D^{(j)}$ starting at the lowest spin $j_o$ appearing. Here any spin $j\geq j_o$ occurs exactly once. See \cite{gelminshap} for the details of this construction. A canonical representation space for $D^{(\rho,j_o)}$ is given by homogenous functions $f:\mathbb{C}^2\rightarrow\mathbb{C},z=(z_o,z_1)\mapsto f(z_o,z_1)$ of degree $(a,b)=(-j_o-1+\I\rho,j_o-1+i\rho)$, an orthogonal basis within these representation spaces is constructed in \cite{unrepsl, lorentzvertam}. Furthermore $\forall\lambda\in\mathbb{C}:f(\lambda z_o,\lambda z_1)=\lambda^a\bar{\lambda}^bf(z_o,z_1)$ defines the degree of homogeneity, and the natural group action is given by right translation, i.e. $(T_gf)(z)=f(zg)$. 

Consider now the real and imaginary parts of the smeared momentum $\Pi_i[f]$ according to the decomposition
\begin{align}
L_i[f] & =  \frac{1}{\hbar}\big(\Pi_i[f]+\bar{\Pi}_i[f]\big),\\
K_i[f] & =  \frac{1}{\I \hbar}\big(\Pi_i[f]-\bar{\Pi}_i[f]\big).
\end{align}
The prefactor of $\hbar$ we've introduced for later convenience only. Notice that these smeared momenta obey the commutation relations of $\mathfrak{sl}(2,\mathbb{C})$ according to:
\begin{align}
\big\{L_i[f],L_j[f^\prime]\big\} & =  -\frac{1}{\hbar}\delta_{ff^\prime}\uo{\epsilon}{ij}{m}L_m[f^\prime],\\
\big\{L_i[f],K_j[f^\prime]\big\} & =  -\frac{1}{\hbar}\delta_{ff^\prime}\uo{\epsilon}{ij}{m}K_m[f^\prime],\\
\big\{K_i[f],K_j[f^\prime]\big\} & =
+\frac{1}{\hbar}\delta_{ff^\prime}\uo{\epsilon}{ij}{m}L_m[f^\prime].
\end{align}
Using this decomposition the densitised triad smeared over some face $f$ turns out to be a sum of both angular momentum and boost components according to
\begin{equation}
\,E_i[f] = \ellp^2\frac{\beta}{\beta^2+1}\left(L_i[f]-\beta K_i[f]\right).
\end{equation}
In the very same manner we can rewrite the smeared version of the reality condition \eqref{ccons} in order to obtain
\begin{equation}
\boxed{C_i[f] = \ellp^2\frac{\beta}{\beta^2+1}\left(K_i[f]+\beta L_i[f]\right)\stackrel{!}{=}0.}
\end{equation}
Where we have apparently recovered the linear simplicity constraints \cite{flppdspinfoam, LQGvertexfinite} of spin-foam gravity.
\subsection{General strategy towards quantisation}
Here as a kind of motivating excursus we would like to briefly sketch the further strategy towards quantum theory. To keep things simple consider Gau\ss\; constraint and reality conditions only. Classically both quantities are demanded to vanish. However in general they do not commute between one another, not even weakly. Instead one finds the following Poisson commutation relations:
\begin{align}
\big\{C_i[f],C_j[f^\prime]\big\}&=
-\frac{\ellp^2}{\hbar}\frac{\beta}{\beta^2+1}\delta_{ff^\prime}\uo{\epsilon}{ij}{l}\big(E_l[f]-\beta C_l[f]\big),\label{ccpoiss}\\
\big\{G_i[\Lambda^i],C_j[f]\big\}&=
-\frac{\Lambda^l-\bar{\Lambda}^l}{2\I}\uo{\epsilon}{ij}{m}E_m[f]
-\frac{\Lambda^l+\bar{\Lambda}^l}{2}\uo{\epsilon}{ij}{m}C_m[f].\label{gcpoiss}
\end{align}
Equation \eqref{gcpoiss} reveals that $C_i[f]$ transforms as a vector under $SU(2)$, but is not constrained to vanish if boosted. Moreover \eqref{ccpoiss} prevents us to implement the reality conditions strongly, that is in the sense of $C_i[f]\Psi=0$. This just follows from the observation that
\begin{equation}
0=\frac{\I}{\hbar}\big[{C}_i[f],{C}_j[f]\big]\Psi=
-\frac{\ellp^2}{\hbar}\frac{\beta}{\beta^2+1}\uo{\epsilon}{ij}{l}E_i[f]\Psi\stackrel{\mbox{in general}}\neq 0
\end{equation}
is a contradiction. Where we simply copied the argument from \cite{dirac}. There are now two independent ways to continue. First one could solve the reality conditions classically. Introducing the Dirac bracket
\begin{equation}
\eqalign{
\big\{F,G\big\}^\star=\big\{F,G\big\}
&-\frac{\hbar}{\ellp^2}\frac{\beta^2+1}{\beta}\int_{p\in\Sigma}\big\{\uo{C}{i}{a}(p),F\big\}\big\{\ou{\Delta}{i}{a}(p),G\big\}+\\
&+\frac{\hbar}{\ellp^2}\frac{\beta^2+1}{\beta}\int_{p\in\Sigma}\big\{\uo{C}{i}{a}(p),G\big\}\big\{\ou{\Delta}{i}{a}(p),F\big\}
}
\end{equation}
one is then led to the reduced phase space $(\mathcal{P}^\star,\{\cdot,\cdot\}^\star)\subset(\mathcal{P},\{\cdot,\cdot\})$ built from the $SU(2)$ Ashtekar connection $\ou{{}^{(\beta)}A}{i}{a}=\ou{\Gamma}{i}{a}+\beta\ou{K}{i}{a}$ and its momentum conjugate $\uo{E}{i}{a}$. The reality conditions together with the vanishing of torsion imply that any point on $\mathcal{P}^\star$ is already determined by the pair $(\ou{{}^{(\beta)}A}{i}{a},\uo{E}{i}{a})$. On the reduced phase space these variables turn out to be canonical conjugate, in fact one obtains the familiar symplectic structure generated by the $SU(2)$ Ashtekar variables, i.e.
\begin{align}
&\big\{\uo{E}{i}{a}(p),\uo{E}{j}{b}(q)\big\}^\star\Big|_{\mathcal{P}^\star}=\big\{\ou{{}^{(\beta)}A}{i}{a}(p),\ou{{}^{(\beta)}A}{j}{b}(q)\big\}^\star\Big|_{\mathcal{P}^\star}=0,\\
&\big\{\uo{E}{i}{a}(p),\ou{{}^{(\beta)}A}{j}{b}(q)\big\}^\star\Big|_{\mathcal{P}^\star}=\frac{\beta\ellp^2}{\hbar}\delta^j_i\delta^a_b\delta^{(3)}(p,q),\\
&\big\{\uo{C}{i}{a}(p),\,\cdot\,\big\}^\star\Big|_{\mathcal{P}^\star}=\big\{\ou{\Delta}{i}{a}(p),\,\cdot\,\big\}^\star\Big|_{\mathcal{P}^\star}=0.
\end{align}
And we would thus recover the theory in its usual Hamiltonian formulation \cite{thiemann, status}, from which the kinematical Hilbert space $\mathcal{K}$ of loop quantum gravity is constructed. 

Inspired by the work of Gupta and Bleuler \cite{gupta, bleuler} one could try something else \cite{flppdspinfoam} though. Suppose we would have found a projector $P$ acting on some yet unspecified kinematical states $\Psi$. Moreover this projector is supposed to annihilate the reality conditions weakly, that is in the sense of
\begin{equation}
PC_i[f]P=0.
\end{equation}
The latter immediately leads to
\begin{equation}
C_i[f]P\Psi\perp P\Psi.\label{weakcons}
\end{equation}
Notice that in order to achieve this orthogonal decomposition no inner product is needed, the symbol $\perp$ refers to nothing but the projector $P$. If we now loosely identify the image of $P$ with the desired kinematical Hilbert space $\mathcal{K}$, we would expect something like
\begin{equation}
\forall\Psi,\Phi\in\mathcal{K}:\langle\Psi,C_i[f]\Phi\rangle=0
\end{equation}
to happen. Hence we would call the reality condition weakly implemented on $\mathcal{K}$. Notice the strong similarity with the classical picture. The Hamiltonian vector field $X_{C_i[f]}=\{C_i[f],\cdot\}$ generates a flow on phase space. But the right hand side of \eqref{ccpoiss} is not demanded to vanish. Therefore this flow actually moves any physical configuration away from the constraint hyper-surface. Something similar would happen in quantum theory too. The kinematical state $P\Psi$ is perpendicular to $C_i[f]P\Psi$, and the operator $C_i[f]$ would thus move any element $P\Psi$ away from $\mathcal{K}$.

What about the Gau\ss\;constraint? Remember first that under the action of the $SU(2)$ subgroup $C_i[f]$ just rotates around some internal axis. Therefore, provided $\Lambda^i$ is real we are allowed to strongly impose the corresponding constraint:
\begin{equation}
\mbox{If}\;\Lambda^i=+\bar{\Lambda}^i:{G}_i[\Lambda^i]P\Psi=0.
\end{equation}
In the case of $\Lambda^i$ being purely imaginary the situation is different. The constraint generates boosts now, and equation \eqref{gcpoiss} would force us to impose it only in the weak sense of: 
\begin{equation}
\mbox{If}\;\Lambda^i=-\bar{\Lambda}^i:{G}_i[\Lambda^i]P\Psi\perp P\Psi.
\end{equation}
The general strategy can now be summarised as follows. Decompose the set of all constraints $\{h_1,\dots,h_N,c_1,\dots,c_M\}$ into two parts. First identify those constraints $h_\mu$ being of first class \cite{dirac, thiemann}, i.e. those $h_\mu$ for which
\begin{equation}
\eqalign{\{h_\mu,h_\nu\}&=\uo{F}{\mu\nu}{\alpha}h_\alpha+\uo{G}{\mu\nu}{i}c_i,\\
\{h_\mu,c_i\}&=\uo{H}{\mu i}{\nu}h_\nu+\uo{I}{\mu i}{j}c_j}
\end{equation}
holds. In quantum theory it is them who can strongly be set to zero, all the other constraints hold weakly, i.e. one searches for some projector $P$ such that
\begin{equation}
\forall \mu: h_\mu P\Psi=0,\quad\mbox{but:}\quad
\forall i: c_iP\Psi\perp P\Psi.
\end{equation}
In the case of the reality conditions this is what we will do in the following.

\subsection{The space of cylindrical functions}
Here, let us try to construct a representation space for the closed algebra (\ref{UUbrace}, \ref{PiUbrace}, \ref{PiPibrace}) of flux and holonomy. To this goal let $f\in\mathcal{C}^\infty(SL(2,\mathbb{C}):\mathbb{C})$ be some smooth function on the group. Let now $g,h\in SL(2,\mathbb{C})$ be group elements and observe that all of the following operators
\begin{align}
&\big(\ou{\widehat{[D^{(\rho,j_o)}]}}{\mu}{\nu}f\big)(g)=\ou{{D^{(\rho,j_o)}(g)}}{\mu}{\nu}f(g),\\
&\big(K_i^\ell f\big)(g)
	=-{\I}\frac{\di}{\di\varepsilon}\Big|_{\varepsilon=0}
	f\big(g\E^{\frac{\sigma_i}{2}\varepsilon}\big),  
\qquad\big(K_i^r f\big)(g)
	=+{\I}\frac{\di}{\di\varepsilon}\Big|_{\varepsilon=0}
	f\big(\E^{\frac{\sigma_i}{2}\varepsilon}g\big),\\
&\big(L_i^\ell f\big)(g)
	=-{\I}\frac{\di}{\di\varepsilon}\Big|_{\varepsilon=0}
	f\big(g\E^{-\frac{\sigma_i}{2\I}\varepsilon}\big), 
\qquad\big(L_i^r f\big)(g)
	=+{\I}\frac{\di}{\di\varepsilon}\Big|_{\varepsilon=0}
	f\big(\E^{-\frac{\sigma_i}{2\I}\varepsilon}g\big),\\
&\big(U_g^\ell f\big)(h)=f(gh),\qquad\big(U_g^r f\big)(h)=f(hg),
\end{align}
actually map the vector space $\mathcal{C}^\infty(SL(2,\mathbb{C}):\mathbb{C})$ onto itself. The labels $\ell$ and $r$ refer to right and left invariance respectively. We have e.g. $(U_g^\ell)_\ast K_i^\ell=K_i^\ell$ and so on. Furthermore $\mu=(j,m)$ with $j=j_o,j_o+1,\dots$ and $m=-j,\dots,j$ is some multi-index referring to the canonical basis within the $(\rho,j_o)$-th unitary irreducible representation space \cite{gelminshap}. On $\mathcal{C}^\infty(SL(2,\mathbb{C}):\mathbb{C})$ one may also wish to introduce the left invariant versions of both the densitised triad and the reality conditions
\begin{align}
E_i^\ell&=\ellp^2\frac{\beta}{\beta^2+1}\big(L_i^\ell-\beta K_i^\ell\big),\\
C_i^\ell&=\ellp^2\frac{\beta}{\beta^2+1}\big(K_i^\ell+\beta L_i^\ell\big).
\end{align}
And analogously for the right invariant part $E_i^r$ and $C_i^r$.
We now choose the function space $\mathcal{C}^\infty(SL(2,\mathbb{C}):\mathbb{C})$ as starting point for the quantisation of the complex theory. To this goal we define the notion of cylindrical functions. This is done in the usual way \cite{rovelli}.

We call a functional $\Psi:\bar{\mathcal{A}}_{SL(2,\mathbb{C})}\rightarrow\mathbb{C}$ mapping any distributional $SL(2,\mathbb{C})$ connection to the complex plane cylindrical, symbolically denoted by $\Psi\in\mathrm{Cyl}$ provided there exists some graph $\Gamma=(\gamma_1,\dots,\gamma_L)$ such that
\begin{equation}
\eqalign{
\Psi\in\mathrm{Cyl}\Leftrightarrow&\exists\Gamma=(\gamma_1,\dots,\gamma_L)\;\mbox{and}\\
&\exists f\in\mathcal{C}^\infty\big(\underbrace{{SL(2,\mathbb{C})}\times\dots\times{SL(2,\mathbb{C})}}_{L\text{-times}}:\mathbb{C}\big)\;\mbox{such that:}\\
&\forall A\in\bar{\mathcal{A}}_{SL(2,\mathbb{C})}:\Psi[A]=f\big(U_{\gamma_1}[A],\dots, U_{\gamma_L}[A]\big).}\label{cyldef}
\end{equation}
On $\mathrm{Cyl}$ the smeared algebra is realised as follows, the matrix elements of the irreducible unitary representations are represented by multiplication operators
\begin{equation}
\big(\ou{\widehat{D^{(\rho,j_o)}(U_\gamma)}}{\mu}{\nu}\Psi\big)[A]  = \ou{{D^{(\rho,j_o)}\big(U_\gamma[A]\big)}}{\mu}{\nu}\Psi[A],
\end{equation}
whereas the momentum variables are introduced as derivatives according to:
\begin{equation}
\eqalign{\qquad
\big(K_i[f_j]\Psi_f)[A] & = -{\I}\frac{\di}{\di\varepsilon}\Big|_{\varepsilon=0}f\big(U_{\gamma_1}[A],\dots,U_{\gamma_j}[A]\E^{\frac{\sigma_i}{2}\varepsilon},\dots,U_{\gamma_L}[A]\big)\\
 &\equiv  \big({\underset{j}{K}}{}^{\ell}_if\big)(U_{\gamma_1}[A],\dots,U_{\gamma_L}[f]).\label{momvar}
}
\end{equation}
Here we only have defined one of the momenta, the quantity $K_i[f^{-1}]$ smeared over the inverted face is represented by $K_i^r$, i.e. the right derivative, and of course the construction for $L_i[f]$ and $L_i[f^{-1}]$ is done in complete analogy. One then soon realises that on $\mathrm{Cyl}$ the Poisson relations (\ref{UUbrace}, \ref{PiUbrace}, \ref{PiPibrace}) are replaced by the corresponding commutation relations, where the Poisson bracket $\{\cdot,\cdot\}$ is replaced by $\I/\hbar$ times the commutator.
\subsection{Solution space for the reality conditions}
In what follows we will recall the Dupuis--Livine map $P:\mathcal{C}^\infty(SL(2,\mathbb{C}):\mathbb{C})\rightarrow\mathcal{C}^\infty(SL(2,\mathbb{C}):\mathbb{C})$ introduced in \cite{liftng, projspinnet} making it possible \cite{lortzcov} to implement the reality conditions weakly, that is in the sense of \eqref{weakcons}. In fact it is their pioneering work that makes this whole construction possible. Following them consider some $f\in\mathcal{C}^\infty(SL(2,\mathbb{C}):\mathbb{C})$ and define for any $g\in SL(2,\mathbb{C})$ the image of $f$ under the action of $P$ according to
\begin{equation}
\eqalign{
\big(Pf\big)(g)=\sum_{2j=0}^\infty(2j+1)&\sum_{m,n=-j}^j\ou{D^{(\beta(j+1),j)}(g)}{(j,m)}{(j,n)}\cdot\\
&\cdot\int_{SU(2)} d\nu_h\ou{\overline{D^{(j)}(h)}}{m}{n}f(h).}\label{dupliv}
\end{equation}
Where $d\nu$ is the Haar measure on the $SU(2)$ subgroup. Both the smoothness of $f$ and the compactness of the group guarantee this integral to be well defined for any $f\in\mathcal{C}^\infty(SL(2,\mathbb{C}):\mathbb{C})$.
It is rather straight forward to show that this map fulfils the following list of elementary properties:
\begin{equation}
\left.\eqalign{
\mbox{(i)}\;& PP=P\\
\mbox{(ii)}\;&\delta^{ij}:\!E_j^\ell C_i^\ell\!: P=\delta^{ij}:\!E_j^r C_i^r\!: P=0\\
\mbox{(iii)}\;&\delta^{ij}:\!C_j^\ell C_i^\ell\!: P=\delta^{ij}:\!C_j^r C_i^r\!: P=0\\
\mbox{(iv)}\;& PC_i^\ell P=PC_i^rP=0\;\forall\,i\in\{1,2,3\}\\
\mbox{(v)}\;&\forall g\in SU(2):U_g^\ell P U_{g^{-1}}^\ell=U_g^r P U_{g^{-1}}^r=P 
}\qquad\qquad\klamm\label{prprts}
\end{equation}
Here :$X$: denotes a suitable ordering \cite{LQGvertexfinite} of the two Casimir operators $C_1=L_iL^i-K_iK^i$ and $C_2=L_iK^i$ of $SL(2,\mathbb{C})$. 

One of the most striking and fascinating properties of this projector is missing in this list. It was Rovelli and Speziale who actually discovered just recently \cite{lortzcov} that the Dupuis--Livine map allows us to implement \emph{local} Lorentz covariance within the spin-foam formalism. What Rovelli and Speziale found is this; when calculating the spin-foam amplitude one assign to each edge some $SU(2)$ subgroup of $SL(2,\mathbb{C})$. But there is no unique choice for this available, in fact to any timelike normal $n^I$ there is a different $SU_n(2)$ subgroup associated. (In our paper by choosing time-gauge $n^I=(1,0,0,0)^I$ we have fixed this subgroup once and for all.) Within the bulk they then prove that the spin-foam amplitude is surprisingly independent of this choice. The only dependence happens to be at the boundary, where any local $SL(2,\mathbb{C})$ transformation can be absorbed by boosting $n^I\mapsto\ou{\Lambda(g)}{I}{J}n^J$ the normal.

Another important observation \cite{lortzcov} is this. From \eqref{dupliv} one immediately finds that the image $Pf$ of $f$ is already entirely determined by the value of $f$ along the $SU(2)$ subgroup. 
Notice also that any element $Pf$ has a natural action on elements $\varphi\in L^2(SL(2,\mathbb{C}),d\mu)$ by means of duality
\begin{equation}
\big(Pf\big)[\varphi]=\int_{SL(2,\mathbb{C})}d\mu_g\overline{\big(Pf\big)(g)}\varphi(g).
\end{equation}
Where $d\mu_g$ denotes the Haar measure 
\begin{equation}
\eqalign{
d\mu_g=\frac{1}{(24\pi)^2\I}\mathrm{Tr}\bigl(g^{-1}\di g&\wedge g^{-1}\di g\wedge g^{-1}\di g\bigr)\wedge\\&\wedge\overline{\mathrm{Tr}\bigl(g^{-1}\di g\wedge g^{-1}\di g\wedge g^{-1}\di g\bigr)}}
\end{equation}
of the $SL(2,\mathbb{C})$ group. However elements of $P\mathcal{C}^\infty(SL(2,\mathbb{C}):\mathbb{C})$ are not normalisable \cite{lortzcov} with respect to the corresponding $L^2$ inner product, but lie within the algebraic dual of $L^2(SL(2,\mathbb{C}),d\mu)$ instead.
Furthermore $P$ can trivially be extended to $\mathcal{C}^\infty(SL(2,\mathbb{C})^L:\mathbb{C})$, which of course we denote by the very same symbol. Similarly we can define the action of $P$ on any element of the space of cylindrical functions by
\begin{equation}
\big(P\Psi_f\big)[A]:=\big(Pf\big)(U_{\gamma_1}[A],\dots,U_{\gamma_L}[A]).\label{PonPsi}
\end{equation}
And again $P$ maps the space $\mathrm{Cyl}$ onto itself. 

Point (iv) of \eqref{prprts} proved by Ding and Rovelli in \cite{physbound} is crucial for what happens next. Their major insight was that within the canonical basis \cite{gelminshap, unrepsl} corresponding to the representation space labelled by $(\rho=\beta(j+1),j_o=j)$ the simplicity constraints hold weakly, that is the combined matrix element
\begin{equation}
\big\langle(\beta(j+1),j),j,m\big|K_i+\beta L_i\big|(\beta(j^\prime+1),j^\prime),j^\prime,m^\prime\big\rangle=0\label{matrxmeth}
\end{equation}
of both boosts and rotations vanishes.

We are now ready to obtain a subspace of $\mathrm{Cyl}$ solving the reality conditions weakly. To this goal we introduce the image of $\mathrm{Cyl}$ under the action of $P$ and call it the space of simple-cylindrical functions denoted by $\mathrm{SCyl}:=P(\mathrm{Cyl})$. Observe that the projector $P$ allows us to decompose $\mathrm{Cyl}$ according to
\begin{equation}
\mathrm{Cyl}=\mathrm{SCyl}\oplus\mathrm{SCyl}^\perp.
\end{equation}
Notice also that for this \qq{orthogonal} decomposition to make sense we do not have to introduce any scalar product on $\mathrm{Cyl}$. Furthermore on $\mathrm{SCyl}$ the reality conditions hold weakly, that is
\begin{equation}
\boxed{
\forall \Psi\in\mathrm{SCyl}\;\mbox{and all faces $f$}:C_i[f]\Psi\perp\mathrm{SCyl}.
}
\end{equation} 
Within this formalism we are now able to reconstruct the kinematical Hilbert space of loop quantum gravity. To do this we first remember that for any $f\in\mathcal{C}^\infty(SL(2,\mathbb{C})^L:\mathbb{C})$ the image $Pf$ is already determined by its restriction to the $SU(2)$ subgroup. Therefore it seems natural to use this restriction in order to define the inner product
\begin{equation}
\langle\Psi,\Phi\rangle=\int_{\bar{\mathcal{A}}_{SU(2)}}d\mu_{\mathrm{AL}}(A)\overline{\big(P\Psi\big)[A]}\big(P\Phi\big)[A]\label{innprod}
\end{equation}
between elements of $\mathrm{Cyl}$. Here $d\mu_{\mathrm{AL}}$ is the Ashtekar--Lewandowski measure on the space of distributional $SU(2)$ connections. However on the space $\mathrm{Cyl}$ this inner product is highly degenerate. Which should not surprise us since $\mathrm{Cyl}$ contains a vast number of unphysical degrees of freedom. To get rid of these extra degrees of freedom one continues as in the Gelfand--Naimark--Segal construction \cite{thiemann}. Introduce the linear subspace
\begin{equation}
\mathcal{I}=\big\{\Psi\in\mathrm{Cyl}\big|\langle\Psi,\Psi\rangle=0\big\}
\end{equation}
in order to define the kinematical Hilbert space $\mathcal{K}$ of loop quantum gravity as the completion of the quotient space associated, i.e.
\begin{equation}
\boxed{\mathcal{K}:=\overline{\mathrm{Cyl}/\mathcal{I}}.}
\end{equation}
We will now show that the subspace of degenerate vectors of the inner product defined as in \eqref{innprod} not only contains $\mathrm{SCyl}^\perp$, but is actually much larger. Consider the following example. Introduce the pair of elements
\begin{equation*}
\eqalign{
\Psi_1[A]&=\ou{D^{(\beta(j+1),j)}(U_{\alpha_2\circ\alpha_1}[A])}{(j,m_1)}{(j,m_o)},\\
\Psi_2[A]&=\sum_{n=-j}^j\ou{D^{(\beta(j+1),j)}(U_{\alpha_2}[A])}{(j,m_1)}{(j,n)}\ou{D^{(\beta(j+1),j)}(U_{\alpha_1}[A])}{(j,n)}{(j,m_o)}}\label{invert}
\end{equation*}
of $\mathrm{SCyl}$ on a single link $\alpha$ divided into sub-links $\alpha_1$ and $\alpha_2$. Here the two sub-links share end and starting points according to $\alpha_2(0)=\alpha_1(1)$. On $\mathrm{Cyl}$ these are two distinct functionals of the $SL(2,\mathbb{C})$ connection, however restricting their argument to $SU(2)$ connections they are perfectly equal. This basically follows from the fact that on the $SU(2)$ subgroup the irreducible unitary representation collapse \cite{gelminshap, lorentzvertam} to the usual $SU(2)$ Wigner matrices according to
\begin{equation}
\forall g\in SU(2):\ou{D^{(\rho,j_o)}(g)}{(j,m)}{(l,n)}=\delta^j_l\ou{D^{(j)}(g)}{m}{n}.
\end{equation}
Therefore it is not hard to see that
\begin{equation}
\langle\Psi_1-\Psi_2,\Psi_1-\Psi_2\rangle=0.
\end{equation}
Thus both $\Psi_1$ and $\Psi_2$ lie within the same equivalence class and must therefore be identified. This can be generalised, instead of splitting $\alpha$ once we can split it $(N-1)$-times, obtaining some $\Psi_N\sim\Psi_1$. Consider now the formal limit of $N\rightarrow\infty$.
\begin{figure}[h]
     \centering
     \includegraphics[width= 0.2\textwidth]{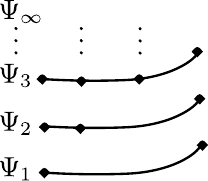}
     \caption{Links distinguishable only by their number of intermediate nodes should be identified.}
     \label{multivert}
\end{figure}
\begin{equation}
{\eqalign{
&\!\!\!\lim_{N\rightarrow\infty}\Psi_N[A]\equiv\lim_{N\rightarrow\infty}\sum_{n_1\dots n_{N-1}=-j}^j \widetilde{D}^{(j)}\Big(\boldsymbol{P}\E^{-\int_{\frac{N-1}{N}}^{1}\di tA_{\gamma(t)}(\dot\gamma)}\Big)^{m_1}_{\phantom{m}n_{N-1}}\cdot\\
&\cdot\widetilde{D}^{(j)}\Big(\boldsymbol{P}\E^{-\int_{\frac{N-2}{N}}^{\frac{N-1}{N}}\di tA_{\gamma(t)}(\dot\gamma)}\Big)^{n_{N-1}}_{\phantom{m}n_{N-2}}\cdots
\widetilde{D}^{(j)}\Big(\boldsymbol{P}\E^{-\int_{0}^{\frac{1}{N}}\di tA_{\gamma(t)}(\dot\gamma)}\Big)^{n_1}_{\phantom{m}m_o}=\\
&\!\!\!=\Big(\boldsymbol{P}\E^{-\int_0^1\di t\widetilde{D}^{(j)}_{\phantom{m}\ast}A_{\gamma(t)}(\dot{\gamma})}\Big)^{m_1}_{\phantom{m}m_o}.}\label{limit}
}
\end{equation}
Where we have introduced the abbreviation
\begin{equation}
\forall g\in SL(2,\mathbb{C}):\ou{\widetilde{D}^{(j)}(g)}{m}{n}
=\ou{D^{(\beta(j+1),j)}(g)}{(j,m)}{(j,n)},
\end{equation}
and $\widetilde{D}^{(j)}_{\phantom{m}\ast}$ is the differential map associated. This however turns out to be equal to:
\begin{equation}
\eqalign{
&\!\!\!\ou{\big(\widetilde{D}^{(j)}_{\phantom{m}\ast}A_{\gamma(t_o)}(\dot{\gamma})\big)}{m}{n}=\\
&=\frac{\di}{\di\varepsilon}\Big|_{\varepsilon=0}\ou{\big[D^{(\beta(j+1),j)}(\boldsymbol{P}\E^{-\int_{t_o}^{t_o+\varepsilon}\di tA_{\gamma(t)}(\dot\gamma)})\big]}{(j,m)}{(j,n)}\\
&=\I\Gamma^i_{\gamma(t_o)}(\dot\gamma)\ou{\big[L^{(\beta(j+1),j)}_i\big]}{(j,m)}{(j,n)} -\I K^i_{\gamma(t_o)}(\dot\gamma)\ou{\big[K^{(\beta(j+1),j)}_i\big]}{(j,m)}{(j,n)}=\\
&=\I\big(\Gamma^i_{\gamma(t_o)}(\dot\gamma)+\beta K^i_{\gamma(t_o)}(\dot\gamma)\big)\ou{\big[L^{(j)}_i\big]}{m}{n}}\label{ash}
\end{equation}
Here $\ou{[L^{(j)}_i]}{m}{n}=\langle j,m|L_i|j,n\rangle$ are the matrix elements of the $SU(2)$ generators in the irreducible spin $j$ representation, and similarly 
\begin{equation*}
\langle(\rho,j_o);j,m|K_i|(\rho,j_o);l,n\rangle=\ou{[K^{(\rho,j_o)}_i]}{(j,m)}{(l,n)}. 
\end{equation*}
In order to obtain the last line of \eqref{ash} we've used the reality conditions in the form of \eqref{matrxmeth} introduced by Ding in \cite{physbound}. Moreover don't confuse here extrinsic curvature $K^i_{\gamma(t)}(\dot\gamma)$ and the generators $K_i^{(\rho,j_o)}$ of $SL(2,\mathbb{C})$. Equation \eqref{ash} strongly suggests to introduce the real valued Ashtekar connection
\begin{equation}
{}^{(\beta)}A=\Gamma+\beta K.
\end{equation}
And we observe that in the limit 
\begin{equation}
\lim_{N\rightarrow\infty}\Psi_N[A]=\ou{D^{(j)}\big(\boldsymbol{P}\E^{-\int_{0}^{1}\di tA^{(\beta)}_{\gamma(t)}(\dot\gamma)}\big)}{m_1}{m_o}\in D^{(j)}\big(SU(2)\big)\label{inflim}
\end{equation}
where infinitely many new nodes are inserted infinitely close to one another, one is naturally led to the holonomy of the $SU(2)$ Ashtekar connection evaluated in the $j$-th irreducible unitary representation of $SU(2)$.
Here $\Psi_N$ is implicitly defined in \eqref{limit}. Notice the vivid reduction of the degrees of freedom here, loosely speaking each $\Psi_N$ depends on $2\times 3\times 3\times\infty$ many real variables ($\ou{A}{i}{a}$ is complex, and $\Psi_N$ depends on both $A$ and $\bar{A}$), but ${}^{(\beta)}\ou{A}{i}{a}$ is real and $\Psi_\infty:=\lim_{N\rightarrow\infty}\Psi_N[A]$ can depend on $3\times 3\times\infty$ many variables \qq{only}. Notice also that this limit coincides with the projected Wilson lines introduced by Sergei Alexandrov in \cite{covhilbert} and together with Livine in \cite{sufromcov}.

The functional $\Psi_\infty[A]$ itself, possessing infinitely many links and nodes, does neither lie within $\mathrm{Cyl}$ nor within the equivalence class associated to $\Psi_1$. Nevertheless this limit does possess a natural home, it can actually be \emph{identified} with the whole equivalence class $[\Psi_1]\in\mathcal{K}=\overline{\mathrm{Cyl}/\mathcal{I}}$. 
Moreover \eqref{inflim} tells us that $\Psi_\infty$ essentially is a cylindrical function of the $SU(2)$ Ashtekar connection, and therefore lies within the original kinematical Hilbert space of loop quantum gravity. It is this identification between $[\Psi_1]$ and the limit $\Psi_\infty$ which---if generalised to arbitrary elements of $\mathcal{K}$---allow us to recover the standard kinematical Hilbert space  \cite{thiemann, status, rovelli} of loop quantum gravity. 

In other words, what has been shown is this; given any $\Psi\in\mathrm{Cyl}$ the corresponding equivalence class $[\Psi]$ contains a sequence $\{\Psi_N\}_{N\in\mathbb{N}}$, where in the limit of $N\rightarrow\infty$ every link is split into infinitely many pieces infinitesimally close to one another. In equation \eqref{limit} this has been done for one particular link. For $N\rightarrow\infty$ the functional $\Psi_\infty[A]=\lim_{N\rightarrow\infty}\Psi_N[A]$ collapses into a proper cylindrical function of the $SU(2)$ Ashtekar connection ${}^{(\beta)}A=\Gamma+\beta K$, and naturally lies within the standard \cite{thiemann, status, rovelli} kinematical Hilbert space $\mathcal{K}_{\mathrm{LQG}}$ of loop quantum gravity. From our elementary definition \eqref{innprod} it then trivially follows that by $[\Psi]\in\mathcal{K}\rightarrow\Psi_\infty\in\mathcal{K}_{\mathrm{LQG}}$ any $[\Psi]$ is isometrically mapped to $\Psi_\infty$. We have thus found an isometry mapping $\mathrm{Cyl}/\mathcal{I}$ to a proper subspace of the standard kinematical Hilbert space  \cite{thiemann, status, rovelli} of loop quantum gravity. 
\section{Conclusion}
\subsection{Summary}
Starting from the Holst Lagrangian we first rewrote the action in terms of complex Ashtekar variables. Switching towards the Hamiltonian formulation we then found additional reality conditions. These reality conditions coincide with the linear simplicity constraints of spin foam gravity. We expressed our hope that this observation may open the possibility to formulate both spin-foam gravity and the canonical formulation of loop quantum gravity on equal footing.

Section \ref{sect4} was dedicated to quantum gravity. After having introduced the classical $SL(2,\mathbb{C})$ holonomy flux algebra, we defined the space $\mathrm{Cyl}$ as its natural carrier space for quantum theory. On $\mathrm{Cyl}$ there is no scalar product available. Therefore $\mathrm{Cyl}$ fails to be a Hilbert space. However in order to impose the simplicity constraints weakly there is no inner product needed. A projector selecting the true kinematical degrees of freedom solving the reality conditions in the sense of \eqref{weakcons} perfectly suffices. The Dupuis--Livine \cite{liftng} map provides a possibility for achieving this. On the resulting subspace a positive inner product was introduced \eqref{innprod}. This inner product is still degenerate, to get rid of this degeneracy elements of $\mathrm{Cyl}$ had to be identified. The corresponding equivalence classes, defining a limit of $SL(2,\mathbb{C})$ spin network functions, can naturally be mapped to the usual kinematical Hilbert space corresponding to the real valued Ashtekar connection ${}^{(\beta)}A=\Gamma+\beta K$.
\subsection{Open issues}
A plenty of questions remain open, the most crucial of them are collected within the following list: 

\emph{(i. Kinematical constraints)} The quantisation of the Gau\ss\;constraint was sketched in section \ref{sect3}. See also \cite{physbound} on that, where it is called closure constraint instead. Rotations are implemented strongly, boosts only weakly. In the case of the vector constraint there is a small subtlety. This constraint generates spatial diffeomorphisms modulo $SL(2,\mathbb{C})$ transformations. In order to implement it strongly, and hence recover the space of $s$-knots, one has to add a term proportional to the Gau\ss\;constraint. The resulting constraint generates diffeomorphisms modulo $SU(2)$ transformations and can then demanded to vanish strongly.

\emph{(ii. Operators on the Hilbert space)} In order to finish the construction of the kinematical Hilbert space we need to talk about operators. Consider, for example the smeared momentum $K_i[f]$ as defined in \eqref{momvar}, i.e. the \qq{boost} part of $\Pi_i[f]$. This operator is perfectly well defined on all of $\mathrm{Cyl}$, but our Hilbert space happens to be a quotient space $\mathcal{K}=\overline{\mathrm{Cyl}/\mathcal{I}}$ now. We may wish to extend this operator to all of $\mathcal{K}$ in the obvious way, that is by defining for any $\Psi\in\mathrm{Cyl}$ and corresponding $[\Psi]\in\mathcal{K}$ that e.g. $K_i[f][\Psi]:=[K_i[f]\Psi]$ (and equivalently for all the other operators on $\mathrm{Cyl}$). Though mathematically reasonable we cannot yet prove this to be the physically right choice. And therefore this certainly deserves further investigations.

\emph{(iii. Torsion)}
Even though an implementation for the reality conditions
 was found, there is still an additional \emph{second class} constraint to fulfil. Torsion must vanish. In the form of \eqref{difftens} this seems impossible to achieve. However \eqref{tordef} does not look so bad. In fact using the volume functional $\boldsymbol{V}$ we could use a kind of Thiemann trick, i.e.
\begin{equation}
D e\propto\{F,\boldsymbol{V}\}
\end{equation}
in order to considerably simplify this constraint. 

\emph{(iv. Dynamics)} It is tempting to expect that beside kinematical similarities dynamics within both spin-foam gravity and the canonical formalism match. According to task 14 of Rovelli's \qq{to do list} \cite{newlook}, does there is a suitable quantisation of the Hamiltonian constraint \eqref{ham} that annihilates the spin foam amplitude
\begin{equation}
Z_{\mathrm{EPRL}}\big(\bar{\Psi}\otimes H[\utilde{N}]\Xi\big)\stackrel{?}{=}0
\end{equation}
available?
\section*{Acknowledgements} 
At the end of this work it is a pleasant duty to thank Prof. Carlo Rovelli, Eugenio Bianchi, and Simone Speziale. They gave me the courage and motivation to follow these lines of thought. I'm grateful for free discussions, clear statements and the open atmosphere they spread. I also want to thank Carlo and Simone for sharing the draft of their current article \cite{lortzcov} before publication. In fact it was precisely this paper cutting some Gordian knot, allowing me to continue with part \ref{sect4} of this article. Simone deserves special thanks for pointing me towards the challenging work of Alexandrov \cite{covhilbert, sufromcov, lqgcov}.


\end{document}